\documentclass[aps, prd, 10pt, twocolumn, superscriptaddress, noshowpacs, preprintnumbers, longbibliography, groupedaddress, footinbib, bibnotes]{revtex4-1}

\usepackage{amsmath}
\usepackage{amsfonts}
\usepackage{amssymb}
\usepackage{bbold}
\usepackage{epsfig}
\usepackage{graphicx}
\usepackage{bm}
\usepackage{array}
\usepackage{hyperref}
\usepackage{listings}
\usepackage{color}
\usepackage[normalem]{ulem}

\newcommand{\fig}{Fig.}

\newcommand{\figu}[1]{\fig~\ref{fig:#1}}

\newcommand{\tf}{f}

\begin{document}

\title{Core-collapse supernovae stymie secret neutrino  interactions}

\author{Shashank Shalgar}
\email{shashank.shalgar@nbi.ku.dk}
\thanks{ORCID: \href{http://orcid.org/0000-0002-2937-6525}{0000-0002-2937-6525}}
\affiliation{Niels Bohr International Academy \& DARK, Niels Bohr Institute,\\University of Copenhagen, 2100 Copenhagen, Denmark}

\author{Irene Tamborra}
\email{tamborra@nbi.ku.dk}
\thanks{ORCID: \href{http://orcid.org/0000-0001-7449-104X}{0000-0001-7449-104X}}
\affiliation{Niels Bohr International Academy \& DARK, Niels Bohr Institute,\\University of Copenhagen, 2100 Copenhagen, Denmark}

\author{Mauricio Bustamante}
\email{mbustamante@nbi.ku.dk}
\thanks{ORCID: \href{http://orcid.org/0000-0001-6923-0865}{0000-0001-6923-0865}}
\affiliation{Niels Bohr International Academy \& DARK, Niels Bohr Institute,\\University of Copenhagen, 2100 Copenhagen, Denmark}

\date{\today}

\begin{abstract}

Beyond-the-Standard-Model interactions of neutrinos among themselves -- {\it secret interactions} -- in the supernova core may prevent the shock revival, halting the supernova explosion.  Besides, if supernova neutrinos en route to Earth undergo secret interactions with relic neutrinos, the  neutrino burst reaching Earth may be down-scattered in energy, falling below the detection threshold.  We  probe secret neutrino interactions through supernova neutrinos and apply our findings to the supernova SN 1987A. We place the most stringent bounds on flavor-universal secret interactions occurring through a new mediator with mass between 10~MeV and 1~GeV.

\end{abstract}

\maketitle

\section{Introduction}\label{sec:intro}

Neutrinos provide a fascinating window into physics beyond the Standard Model.  In particular, well-motivated extensions of the Standard Model posit the existence of new {\it secret neutrino interactions} ($\nu$SI). Secret interactions may lead to significant enhancements to the otherwise feeble neutrino-neutrino interactions, and have a rich phenomenology.

\begin{figure}[t!]
 \centering
 \includegraphics[width=\columnwidth,height=\columnwidth]{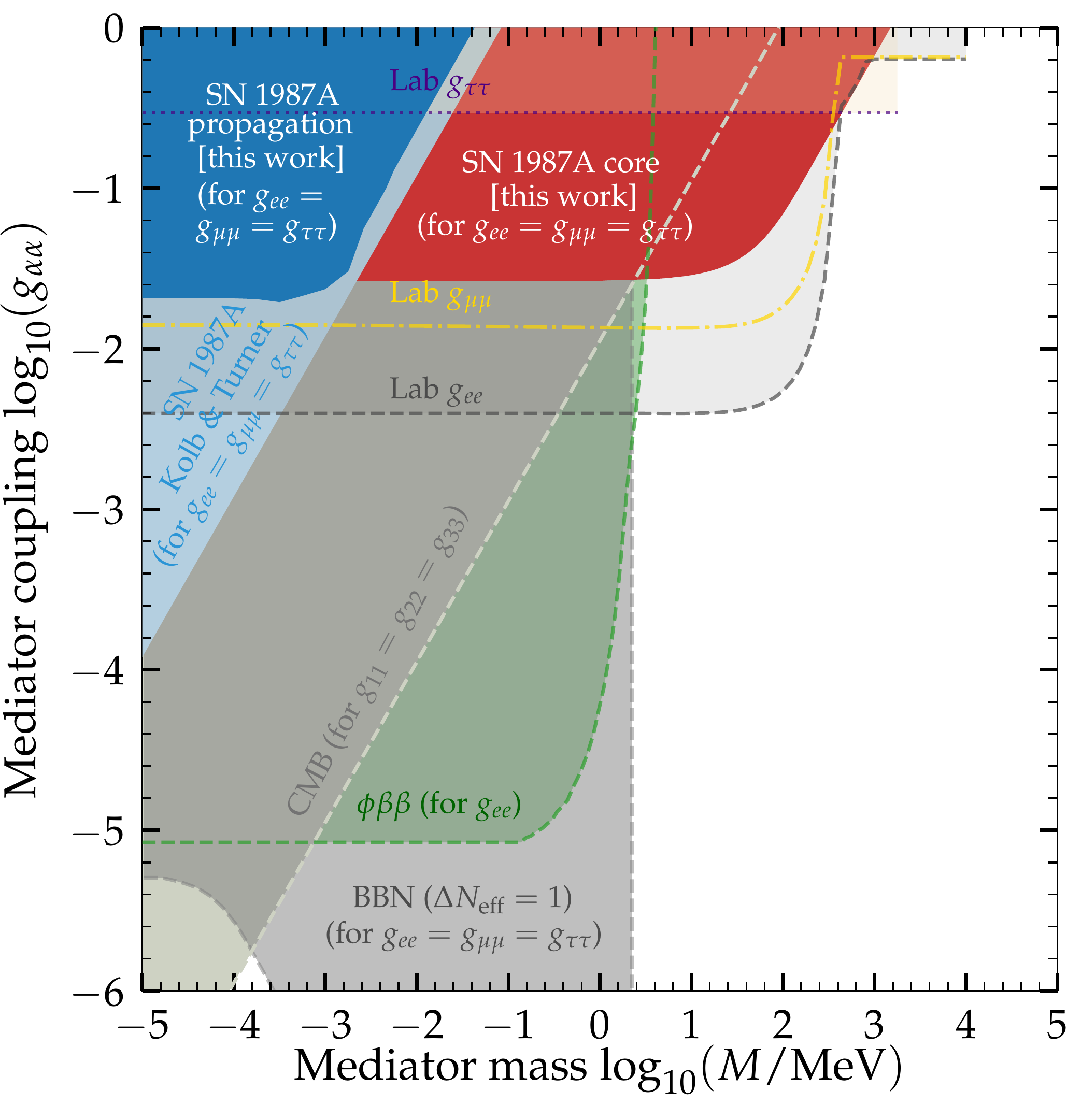}
 \caption{\label{fig:fig1}Constraints on secret neutrino interactions ($\nu$SI), in terms of the coupling $g$ and mass $M$ of the new $\nu$SI mediator.  Our new constraints come from considering $\nu$SI between SN neutrinos en route to Earth and C$\nu$B neutrinos (``SN 1987A propagation'') and $\nu$SI between neutrinos in the SN core (``SN 1987A core''). We consider $\nu$SIs that are flavor-universal ($g_{ee}=g_{\mu\mu}=g_{\tau\tau}$).  Because of this, for the SN1987A core bound, although electron-type neutrinos interact the most in the SN core, our bound applies to all neutrino flavors equally. An earlier SN constraint (``SN 1987A Kolb \& Turner'')~\cite{Kolb:1987qy} comes from the strength of the $\nu$SI interaction rate of neutrinos from the SN 1987A en route to Earth, but our refined treatment supersedes it.  Other constraints come from Big Bang Nucleosynthesis (BBN for $g_{ee}=g_{\mu\mu}=g_{\tau\tau}$)~\cite{Huang:2017egl}, particle decays (we distinguish between flavors, ``Lab $g_{ee}$'', ``Lab $g_{\mu\mu}$'', ``Lab $g_{\tau\tau}$''~\cite{Berryman:2018ogk})~\cite{Laha:2013xua}, double beta decay ($\phi\beta\beta$, {\color{red} for $g_{ee}$})~\cite{Brune:2018sab}, and the cosmic microwave background (CMB for $g_{11}=g_{22}=g_{33}$, in the neutrino mass eigenstate basis)~\cite{Archidiacono:2013dua}.  For cosmological bounds, see also Ref.~\cite{Blinov:2019gcj}.
}
\end{figure}

Secret neutrino interactions occur via a new mediator that couples to neutrinos.  Its mass $M$ and coupling strength $g$ are not known a priori.  
Presently, there is no evidence for $\nu$SI, but there is a wide variety of $\nu$SI models, motivated as solutions to open issues, including the origin of neutrino 
mass~\cite{Chikashige:1980ui, Gelmini:1980re, Georgi:1981pg, Gelmini:1982rr, Nussinov:1982wu, Blum:2014ewa}, tensions  in cosmology~\cite{Aarssen:2012fx, Cherry:2014xra, Barenboim:2019tux}, the muon anomalous moment~\cite{Araki:2014ona, Araki:2015mya}, and the LSND anomaly~\cite{Jones:2019tow}.
Constraints on $\nu$SI come from particle physics, cosmology, and astrophysics, as shown in \figu{fig1}. 

In particle physics, the decay width of particles whose final state contains neutrinos can be affected by $\nu$SI.  The weak decays of the $W$ boson and the neutral $K$ meson have been used to exclude $M < \mathcal{O}(10)$~MeV and $g \gtrsim 10^{-9}$~\cite{Laha:2013xua}.

In cosmology, if the $\nu$SI mediator thermalizes in the early Universe, it introduces additional degrees of freedom that contribute to the total entropy.  This scenario is constrained by the Big Bang Nucleosynthesis (BBN) yields and excludes $M < \mathcal{O}(1)$~MeV and $g \gtrsim 10^{-10}$~\cite{Ahlgren:2013wba, Huang:2017egl,Blinov:2019gcj}.  Separately, $\nu$SI are constrained by observations of the cosmic microwave background (CMB).  Cosmic microwave background anisotropies depend on the anisotropy of the neutrino field  strongly.  Secret neutrino interactions would isotropize the neutrino field, affecting the CMB.  This argument excludes $M < \mathcal{O}(1)$~MeV and $g \gtrsim 10^{-7}$~\cite{Archidiacono:2013dua}.

In astrophysics, neutrinos provide independent means to test for $\nu$SI.
Secret interactions may affect neutrino self-interactions within the astrophysical source itself, if the neutrino density is high enough, like in core-collapse supernovae (SNe), or induce an elastic scattering of astrophysical neutrinos off the cosmic neutrino background (C$\nu$B) as they propagate to Earth.  Astrophysical neutrinos have the potential to probe $\nu$SI with mediator mass  up to $M \gtrsim \mathcal{O}(1)$~GeV, i.e., they can probe mediator masses significantly higher than other existing probes.  

Pioneering work from Ref.~\cite{Kolb:1987qy} invoked the observation of MeV neutrinos from SN 1987A to constrain the effect of $\nu$SI between SN neutrinos propagating to Earth and the C$\nu$B.  
References~\cite{Ioka:2014kca, Ng:2014pca, Blum:2014ewa, Ibe:2014pja, Farzan:2014gza, DiFranzo:2015qea, Kelly:2018tyg, Murase:2019xqi} showed that the occurrence of $\nu$SI between high-energy astrophysical neutrinos and the C$\nu$B would distort the energy spectrum of the astrophysical neutrinos by introducing a deficit at high energies and a pile-up at low energies.  Recently, Ref.~\cite{Murase:2019xqi} studied the potential delay in the arrival times at Earth of TeV--PeV neutrinos as a result of their scattering off the C$\nu$B via $\nu$SI;  if a source emits a burst of high-energy neutrinos and gamma rays simultaneously, neutrinos that undergo $\nu$SI on their way to Earth would take a longer path and arrive later than gamma rays.

In this work, we use MeV neutrinos from Galactic core-collapse SNe to place constraints on $\nu$SI. 
Throughout, we take the prevalent view that the SN explosion is powered by the neutrino-driven mechanism, and that the SN neutrino emission is significant~\cite{2017hsn..book.1095J,Burrows:2012ew}.
Severe $\nu$SI would invalidate both statements.  We investigate two scenarios:  $\nu$SI in the SN core and during the propagation of neutrinos to Earth.  We frame these two scenarios in the context of the neutrinos detected from SN 1987A.

In the SN core, the neutrino density is high enough for neutrinos to be trapped, a situation that favors the potential occurrence of $\nu$SI at an important rate. To place our constraints, we consider the next-to-leading order $\nu + \bar{\nu} \rightarrow 2 \nu + 2 \bar{\nu}$, $\nu + \nu \rightarrow 3 \nu + \bar{\nu}$ and $\bar{\nu} + \bar{\nu} \rightarrow \nu + 3 \bar{\nu}$ (collectively denoted by $2\nu \rightarrow 4\nu$ henceforth).  Conservatively, if each neutrino in the SN core undergoes this process once, the neutrino number density is doubled while the neutrino average energy is halved.  As a result, neutrinos may be unable to transfer enough energy to the stalled SN shock wave to revive it,  halting the explosion.  

During their propagation to Earth, SN neutrinos may scatter off the C$\nu$B, predominantly via the leading-order $\nu$SI process $\nu + \bar{\nu} \rightarrow \nu + \bar{\nu}$.  Supernova neutrinos would suffer a severe shift towards low energies, potentially falling below the  energy threshold for detection, of around 5~MeV.  Further, the scattering would deflect neutrinos from their original propagation direction, significantly delaying their arrival time at Earth.  We treat both effects jointly during the propagation, refining and extending the treatment from Ref.~\cite{Kolb:1987qy}.  In view of the complexity of the neutrino-driven SN mechanism, our goal is to find limits on $\nu$SI that hinge only on the general features of the SN neutrino emission properties.

Figure \ref{fig:fig1} shows our results.  Constraints from $\nu$SI in the SN core disfavor $g \gtrsim 10^{-1.8}$ and $M \lesssim 15$~GeV.  Constraints from $\nu$SI during propagation are weaker and apply only to $M \lesssim 25$~keV.  Between 10~MeV and 15~GeV, our constraints are the strongest to date.  
Since $2 \nu \rightarrow 4 \nu$ is a next-to-leading-order process, one may be inclined to believe that $\nu$SI in the core would lead to negligible effects compared to $\nu$SI during propagation.  Surprisingly, this is not the case; below we explain why.
 
This paper is organized as follows. Section~\ref{sec:nuSI} introduces $\nu$SI.  Section~\ref{sec:core} discusses  the effect of $\nu$SI in the SN core.  Section~\ref{sec:propagation} discusses the effect of $\nu$SI during propagation.  Section~\ref{sec:conclusion} summarizes and concludes.


\section{Secret neutrino interactions}\label{sec:nuSI}  

In order to constrain a large class of $\nu$SI models, we adopt an effective-field-theory approach rather than focus on specific models.  As a result, our limits on $\nu$SI are of wide applicability, but must be interpreted carefully: at neutrino energies well above the scale of the mediator mass, computing the neutrino-neutrino scattering amplitude precisely would require abandoning our effective-field-theory approach and adopting a specific $\nu$SI model. 

The mediator of $\nu$SI can be massless~\cite{Gelmini:1980re, Georgi:1981pg, Gelmini:1982rr, Nussinov:1982wu, Blum:2014ewa}, such as the Majoron; heavy~\cite{Kolb:1987qy, Bilenky:1992xn, Bilenky:1994ma, Bilenky:1999dn}, and treated via an effective field theory; or of intermediate mass~\cite{Chacko:2003dt, Chacko:2004cz, Davoudiasl:2005ks, Goldberg:2005yw, Baker:2006gm}, and introduce resonances. We focus on intermediate-mass mediators because they may introduce detectable imprints on astrophysical neutrinos.  

The $\nu$SI mediator can be a scalar (or pseudo-scalar) or vector (or axial-vector) boson~\cite{Kelly:2018tyg}.  In what follows, we adopt a scalar, $\phi$, yet the limits on $\nu$SI that we place are valid for a scalar and a vector mediator, as we explain below.  The $\nu$SI interaction is described by
\begin{equation}
 \mathcal{L} = g_{\alpha \beta} \bar{\nu}_\alpha \nu_\beta \phi \ ,
\end{equation} 
where $\alpha, \beta = e, \mu, \tau$.  For simplicity, we assume that the interaction is diagonal and universal, i.e., that the only non-zero entries are $g_{\alpha\alpha} \equiv g$, so that all flavors of neutrinos and antineutrinos are affected equally. Since the electron-type neutrinos play a dominant role in the neutrino-driven explosion mechanism, our results are valid as long as the electron-type neutrinos participate in the $\nu$SI. However, in general, it possible to have $\nu$SIs which predominantly affect non-electron type neutrinos, in which case our approach is not applicable. Because the mediator is a scalar, its decay $\phi \rightarrow \nu + \bar{\nu}$ is helicity-suppressed.  Hence, the limit that we obtain should be interpreted as limit on an effective coupling that includes the helicity-suppression factor. The helicity-suppression argument only applies  in the case of a lepton-number-conserving coupling, which is not possible in the case of Dirac neutrinos. In the case of Majorana neutrinos, the $\nu$SI coupling may or may not be the source of lepton number violation. For simplicty, we assume that the $\nu$SI coupling is lepton-number-conserving for the case of a scalar mediator. When interpreting our limits for a vector mediator, there is no such helicity-suppression factor; hence, the limits apply directly on the coupling to the vector mediator. 

In the SN core, we test $\nu$SI through the next-to-leading-order process $2 \nu \rightarrow 4 \nu$, where both interacting neutrinos are SN neutrinos with energies of up to approximately 100~MeV.  We estimate the cross section $\sigma_{2\nu \to 4\nu}$ using dimensional analysis, which is a sufficient approximation for our purposes.  Each final-state particle introduces a factor of $(2\pi)^{-3} \sim 10^{-2}$~\cite{[][{ Page 402}]Tait:2013toa}, arising from the phase-space factor associated to each particle. For center-mass-energies below the mediator mass,
\begin{equation}
 \sigma_{2\nu \to 4\nu}(E_{\rm{CM}}) = \left(10^{-2}\right)^{4} \frac{g^{8}E_{\rm{CM}}^{6}}{M^{8}}\ \quad \textrm{for } E_{\rm{CM}} < M\ ,
 \label{equ:sigone}
\end{equation}
where $E_{\rm{CM}}$ is the total center-of-mass energy.  For center-of-mass energies above the mediator mass, the cross section is independent of the mediator mass, 
\begin{equation}
 \sigma_{2\nu \to 4\nu}(E_{\rm{CM}}) = \left(10^{-2}\right)^{4} g^{8}E_{\rm{CM}}^{2}\ \quad \textrm{for } E_{\rm{CM}} > M\ .
 \label{equ:sigtwo}
\end{equation} 

We have verified that the energy dependence of $\sigma_{2\nu \to 4\nu}$ above is consistent with what we obtain by numerically computing the cross section from one of the contributing Feynman diagrams using CalcHEP~\cite{Belyaev:2012qa} (see the Appendix).  There is a broad, small resonance between $M$ and $2M$, but it does not affect the overall results.  Because the cross section depends strongly on $g$ (i.e., $\sigma_{2\nu \to 4\nu} \varpropto g^8$), our results are fairly independent of small correction factors omitted in our approximations.  Moreover, we have been conservative in our estimation: due to the large number of contributing diagrams, the cross section, when computed in full, may be larger.  Hence, the constraints that we derive from $\nu$SI in the SN core are conservative.

During the propagation of SN neutrinos to Earth, we test $\nu$SI through the leading-order process $\nu + \bar{\nu} \to \nu + \bar{\nu}$, where one of the interacting neutrinos is a SN neutrino with an energy of $\mathcal{O}(10)$~MeV and the other is a C$\nu$B neutrino with $\mathcal{O}(0.1)$~meV energy. 
Compared to SN neutrinos, neutrinos in the C$\nu$B are essentially at rest.  
Thus, when an incoming SN neutrino scatters off the C$\nu$B, the deflection angle of the outgoing relativistic neutrino is completely determined by the 
energies of the incoming and outgoing neutrinos.
The distribution of the outgoing directions is given by the differential cross section for the $s$-channel process, which we take to have a Breit-Wigner form (see, e.g., Refs.~\cite{Ioka:2014kca, Ng:2014pca, Blum:2014ewa, Farzan:2014gza, DiFranzo:2015qea, Kelly:2018tyg}), i.e.,
\begin{equation}
 \frac{d\sigma_{2\nu \to 2 \nu}(E_i, \tilde\theta)}{d\cos\tilde\theta}
 =
 \frac{1}{8\pi}
 \frac{g^4 s^2}{(s-M^2)^2+M^2\Gamma^2} 
 \left(\frac{E_f(E_i,\tilde\theta)}{m_\nu E_i}\right)^2 \ ,
 \label{equ:sigma}
\end{equation}
in the lab frame.  Here, $E_i$ is the initial energy of the SN neutrino, $E_f(E_i,\tilde{\theta}) \equiv (s / 2) [E_i (1-\cos \tilde\theta) + m_\nu]^{-1}$ is the final energy of the SN neutrino, $\tilde\theta \equiv \theta_f-\theta_i$ is the angle between the initial ($\theta_i$) and final ($\theta_f$) directions of the SN neutrino, $\sqrt{s} \equiv \sqrt{2 E_i m_\nu}$ is the center-of-mass energy, $m_\nu$ is the neutrino mass, and $\Gamma \equiv g^2 M / (4\pi)$ is the decay width of the mediator.  
It should be noted that for a given intitial energy, the final energy is uniquely determined by $\tilde{\theta}$ and, consequently, it is possible to write the same differential cross section in terms of the final energy,
\begin{equation}
\frac{d\sigma_{2\nu \to 2 \nu}(E_i, \tilde\theta)}{d\cos\tilde\theta} = \frac{E_{f}^{2}}{m_{\nu}} \frac{d\sigma_{2\nu \to 2 \nu}(E_i, \tilde\theta)}{d E_{f}}.
\label{jac}
\end{equation}

The differential cross section can thus be either interpreted as giving us the probability of deflection by a certain angle, or equivalently, the probability that the energy will change from $E_{i}$ to $E_{f}$.

The SN neutrino is highly relativistic, the scattering is very forward-peaked, so $\tilde\theta$ is small; this becomes relevant later, when propagating SN neutrinos to Earth.  The cross section is resonant when the neutrino energy is $E_{\rm res} = M^2/(2 m_\nu)$.  In comparison, the cross section for the $t$-channel process is small~\cite{Ng:2014pca, Kelly:2018tyg}, and we neglect it here.  To produce our results, we fix $m_\nu = 0.05$~eV, in agreement with the most recent bounds on the sum of neutrino masses, from cosmology~\cite{Capozzi:2018ubv, RoyChoudhury:2019hls}. 

The distribution of neutrino scattering angles is different for scalar and vector $\nu$SI mediators. As a consequence, the differential cross section in Eq.~\ref{equ:sigma} should be modified by a constant factor of $\mathcal{O}(1)$ for vector $\nu$SI mediators. However, we neglect this small correction since  our bounds, as we show later, purely rely  on the reduction in the neutrino energy induced by the $\nu$SI, which is common to both types of mediator. As such, our bounds on $\nu$SI are independent of whether the mediator is a scalar or a vector. Added to that, inside the SN core, because the interacting neutrinos are thermalized and uniformly distributed, the directions of the final-state neutrinos that emerge from a $\nu$SI scattering are also uniformly distributed.  Further, during propagation to Earth, the final-state relativistic neutrino that emerges from a $\nu$SI interaction is highly boosted in the forward direction, so any differences in the angular distribution between a scalar and a vector mediator are sub-dominant.


\section{Secret neutrino interactions in the supernova core}\label{sec:core} 

Because  SNe are extremely dense in neutrinos, with typical neutrino densities of $n_\nu \sim \mathcal{O}(10^{38})$~cm$^{-3}$ deep in the core of the proto-neutron star~\cite{2017hsn..book.1095J,Mirizzi:2015eza}, $\nu$SI can be significant and affect both the neutrino emission and the SN explosion mechanism itself.  Below, we present simple yet powerful arguments to identify the regions in the $(g, M)$ parameter space where $\nu$SI would significantly affect the SN explosion by relying on the $\nu$SI process $\nu + \bar{\nu} \rightarrow 2\nu + 2\bar{\nu}$.  The  robustness of our argument stems from the relative independence on the precise values of the SN input parameters that we choose.

Inside the dense proto-neutron star, neutrinos are trapped due to their frequent interactions with nucleons.  Matter falling onto the  core bounces off of it, creating a shock wave that expands outwards, but that is soon stalled.  Within roughly 500~ms after the bounce, neutrinos that  escape from the SN core are thought to play a vital role in reviving the stalled shock wave and triggering the explosion~\cite{2017hsn..book.1095J}.  Neutrinos do so by depositing energy though scatterings with the medium.  This is the so-called ``neutrino-delayed explosion mechanism''~\cite{Bethe:1984ux}.  

The amount of energy deposited by neutrinos to revive the shock is approximately
\begin{equation}
 E_{\rm dep} \propto n_\nu \sigma_{\nu N} \propto n_\nu \langle E \rangle^{2}\ ,
 \label{equ:Edep}
\end{equation}
where $n_{\nu}$ is the neutrino number density summed over all six flavors, $\langle E \rangle$ is the average neutrino energy, and $\sigma_{\nu N} \propto \langle E \rangle^2$ is the neutrino-nucleon cross section.  The amount of energy deposited by neutrinos depends on the flavor composition, since electron-type neutrinos have a larger cross section than non-electron ones, but, for the sake of simplicity, we ignore this effect.  Further, we assume zero chemical potential for all flavors in the core, since we focus on SN regions where the degeneracy parameter is smaller than unity. 

\begin{figure}[t!]
 \centering
 \includegraphics[width=\columnwidth]{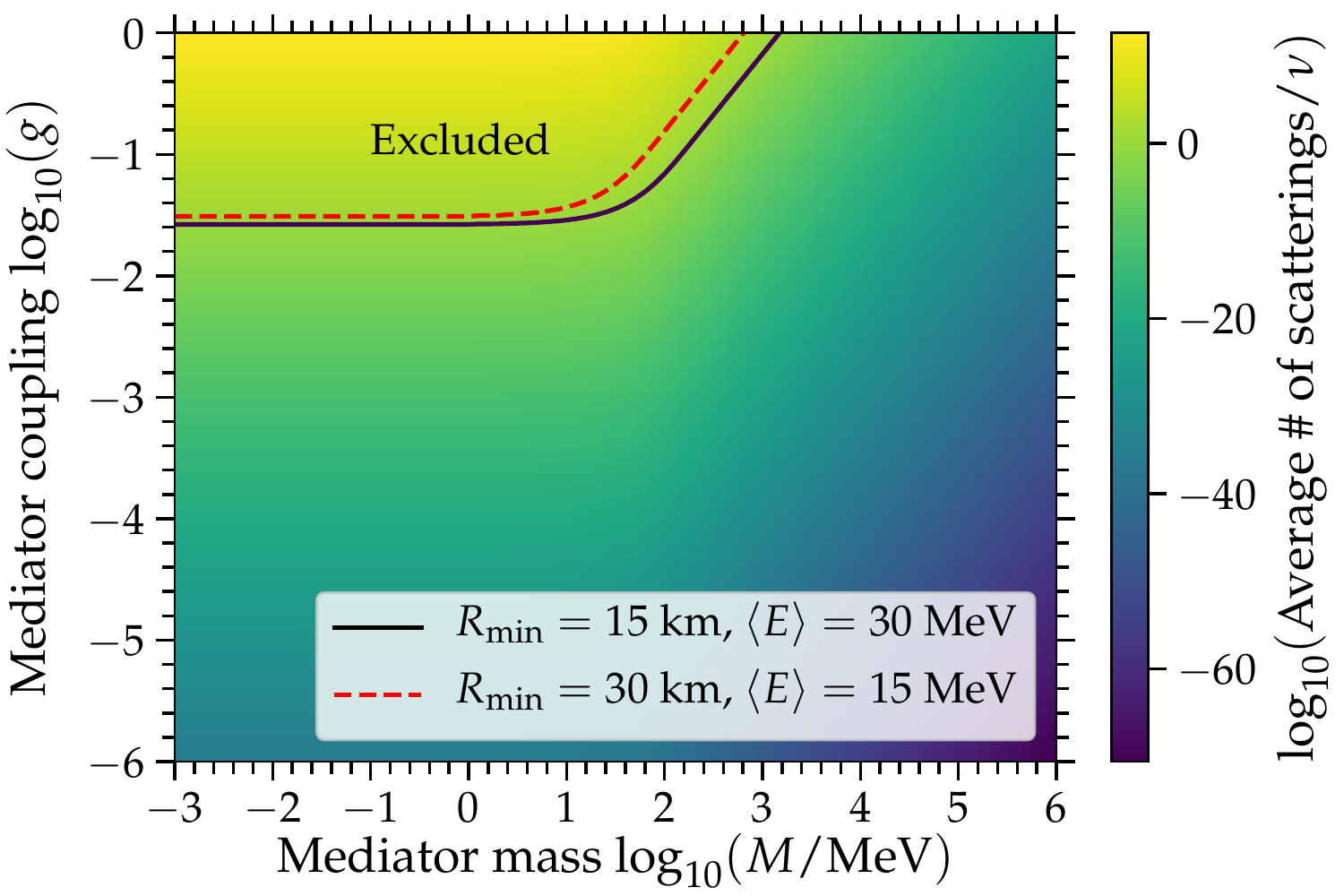}
 \caption{\label{fig:fig2} Constraints on the mass and coupling of the $\nu$SI mediator based on $\nu$SI  occurring in the SN core.  The color coding represents the average number $N_{2\nu \rightarrow 4\nu}$ of $\nu$SI $\nu + \bar{\nu} \rightarrow 2 \nu + 2 \bar{\nu}$ interactions that occur in the SN, see Eq.~\ref{equ:numcoll}.  The region above the black line is excluded for the default case with $R_{\textrm{min}}=15$~km and $\langle E \rangle = 30$~MeV. The region above the dashed line is the exclusion region for the consevative case with $R_{\textrm{min}}=30$~km $\langle E \rangle = 15$~MeV. There, $N_{2\nu \rightarrow 4\nu} \geq 1$ and the neutrino energy deposition is insufficient to revive the stalled shock wave, thus halting the SN explosion.
}
\end{figure}

Because the leading-order $\nu$SI process $\nu + \bar{\nu} \rightarrow \nu + \bar{\nu}$ preserves the number of neutrinos and the average neutrino energy, it does not affect $E_{\rm dep}$.  
In contrast, the next-order processes $2 \nu \rightarrow 4 \nu$ increases the number of neutrinos and decreases the average neutrino energy, i.e., it changes $n_{\nu} \rightarrow n^\prime_{\nu} \simeq \alpha n_{\nu}$ and $\langle E \rangle \rightarrow \langle E^\prime\rangle \simeq \langle E\rangle/\alpha$, where $\alpha > 1$ is a constant factor that depends on the $\nu$SI interaction rate in the core.  As a result, the deposited energy (Eq.~\ref{equ:Edep}) decreases by a factor $\alpha$: $E_{\rm dep} \rightarrow E_{\rm dep}^\prime = n_\nu^\prime \sigma_{\nu N}^\prime \simeq E_{\rm dep}/\alpha$~\footnote{Since different neutrino flavors deposit energy at a different rate, flavor conversions occurring near the neutrino decoupling region, i.e., the region where the neutrino optical depth is low enough that neutrinos can free-stream, may also affect $E_{\rm dep}$, but this effect is negligible compared to the one from $\nu$SI.}.  Large values of $\alpha$ mean that $\nu$SI in the core would render $E^\prime_{\rm dep}$ too small to revive the SN shock, halting the explosion.  Below, we estimate the values of $M$ and $g$ for which this would occur.  

We make the reasonable assumption that the dominant Standard-Model interaction rate in the SN core is due to neutrino-nucleon scattering.  This process determines the radius $R_\nu$ of the ``neutrino-sphere,'' the roughly spherical region within which neutrinos are trapped.  In the early stages of the SN, $R_\nu$ is of a few tens of km, and decreases with time. 
Inside the neutrino-sphere, neutrinos are thermalized with the nucleons of the medium and their number density $n_\nu$ follows a Fermi-Dirac distribution, with temperature $T \simeq \langle E \rangle / 3.15$.  Notably, deep inside within the neutrino-sphere neutrinos are degenerate (i.e., Pauli blocking is high), so that there is no room for their number to grow via the $\nu$SI process $2 \nu \rightarrow 4 \nu$.
Thus, we focus instead on the ``neutrino decoupling region'', i.e., a region in the proximity of the neutrino-sphere where the number density of nucleons starts falling and neutrinos gradually decouple from nucleons.  There, the neutrino density is still high, but the neutrino degeneracy is negligible.  These two conditions allow for the process $2 \nu \rightarrow 4 \nu$ to potentially occur at a high rate.  We assume that the neutrino number density still follows a Fermi-Dirac distribution in the decoupling region, and we fix the average energy to $\langle E \rangle = 30$~MeV, a value representative of expectations;~see, e.g., Ref.~\cite{Tamborra:2017ubu}.

In the decoupling region, neutrinos move along a random walk as they scatter off residual nucleons.
Their energy-averaged mean free path is $\lambda_{\nu N} = \left[ \int dE (dn_\nu(E)/dE) \sigma_{\nu N}(E) \right]^{-1}$.  Hence, the total number of neutrino-nucleon scatterings that a neutrino undergoes is, on average, $N_{\nu N} = \int_{R_{\rm{min}}}^{R_{\rm{max}}} dr\,  r/\lambda_{\nu N}^2$, according to the central limit theorem~\cite{Chandrasekhar:1943ws}.  Here, $R_{\rm{min}} = 15$~km is the approximate, time-averaged minimum radius where the neutrino degeneracy starts to be negligible and  decoupling begins, while $R_{\rm{max}}$ is the maximum radius of the decoupling region.  Because the neutrino and nucleon densities fall steeply with radius, $N_{\nu N}$ is dominated by the interactions that occur closest to the neutrino-sphere.  As a result, the exact choice for $R_{\rm{max}}$ does not matter; we set it to $45$~km.  After $N_{\nu N}$ scatterings, a neutrino has traveled a path of length $d \equiv \lambda_{\nu N} N_{\nu N}$, on average.

If $\nu$SI  occur during the neutrino random walk, the average number of $2\nu \rightarrow 4 \nu$ scatterings that a neutrino undergoes is
\begin{equation}
 N_{2\nu \rightarrow 4\nu} 
 \simeq
 \sigma_{2\nu \to 4\nu} n_{\nu} d 
 \simeq
 \sigma_{2\nu \to 4\nu} n_{\nu} \lambda_{\nu N} N_{\nu N}\ .
 \label{equ:numcoll}
\end{equation}
We assume that, in the decoupling region, neutrinos have a fairly isotropic distribution, and hence the typical $\nu$SI center-of-mass energy is $2 \langle E \rangle$.  Neutrinos are produced at different locations in the SN core and propagate along different paths while depositing energy to revive the shock.   

Conservatively, if $N_{2\nu \rightarrow 4\nu} = 1$, a neutrino undergoes a single $\nu$SI scattering, on average.  In this case, $\alpha = 2$ and this implies $n^\prime_\nu = 2 n_\nu$ and $\langle E^\prime \rangle = \langle E \rangle/2$.  
According to Eq.~\ref{equ:Edep}, this is sufficient to reduce the energy  deposited by neutrinos to revive the shock  by $50\%$ with respect to the case without $\nu$SI and, therefore, halt the SN explosion. 

The dependence of $N_{2\nu \rightarrow 4\nu}$ on $g$ and $M$ is very strong and, consequently, the limits on $g$ and $M$ are very robust to some of the assumptions that we have made.  We illustrate this by comparing the limits obtained using our default case with $R_{\textrm{min}}=15$~km and $\langle E \rangle = 30$~MeV against an alternative, more conservative, case with $R_{\textrm{min}}=30$~km and $\langle E \rangle = 15$~MeV.  As long as we maintain the criterion of $N_{2\nu \rightarrow 4\nu} = 1$ for placing limits, the latter are practically unchanged; this is clearly visible in Fig.~\ref{fig:fig2}, where the limit obtained in the default case is close to the limit obtained in the alternative case, which is slightly more conservative.

If $N_{2\nu \rightarrow 4\nu} \gg 1$, the average neutrino energy decreases progressively with each $\nu$SI scattering.  Therefore, the probability of neutrinos interacting with each other may change over time.  We do not consider this feedback effect, since our constraints on $\nu$SI are already strong even considering that neutrinos interact only once.

Figure~\ref{fig:fig2} shows the region of the $(M,g)$ parameter space excluded by requiring $N_{2\nu \rightarrow 4\nu} \ge 1$ in the SN core. The cross section for $2 \nu \rightarrow 4 \nu $ explains the shape of the constrained region.  For masses $M \lesssim 100$~MeV, the cross section (Eq.~\ref{equ:sigtwo}) is independent of $M$, and so is our constraint.  For masses $M \gtrsim 100$~MeV, the cross section (Eq.~\ref{equ:sigone}) depends on $M$, and our constraint weakens with rising values of $M$.  Later, we find a similar behavior for the constraints based on the propagation of SN neutrinos to Earth.

Our results are based solely on the fact that $\nu$SI reduce the average neutrino energy.  As a consequence, our constraint is only weakly dependent on the exact SN inputs, such as the temperature of the medium, and, therefore, applies to the observation of neutrinos from SN 1987A, under the assumption that its explosion occurred via the neutrino-delayed mechanism.  
Figure~\ref{fig:fig1} shows our constraint from inside the SN core (``SN 1987A core'').  For masses from $M \approx 10$~MeV to 15~GeV, our bound is the strongest and, for $M \gtrsim 100$~MeV, it is the first.

There is one subtle point in our approach.  It concerns the occurrence of the reverse reaction, $4 \nu \rightarrow 2\nu$.  At high densities, deep in the SN core, the dynamics of the forward and reverse reactions is complex: either the forward reaction is blocked due to the high neutrino degeneracy or the forward and reverse reactions are in equilibrium.  We have bypassed this issue by focusing instead on the decoupling region and beyond.
There, we assume that the forward and reverse reactions are not in equilibrium anymore, since the $4\nu \rightarrow 2\nu$ process is disfavored by the decreasing neutrino number density and average energy.


\section{Secret interactions of supernova neutrinos en route to Earth}\label{sec:propagation}  

Supernova neutrinos en route to Earth may scatter on the C$\nu$B via $\nu$SI, predominantly through $2 \nu \rightarrow 2 \nu $.  As a result, their energies, directions, and arrival times at Earth would be affected.  Below, we show that if these effects are severe, SN neutrinos would become undetectable, and we find the values of $M$ and $g$ for which this happens.

Supernova neutrinos are affected by $\nu$SI in two ways.  First, given the lower energy of C$\nu$B neutrinos versus SN neutrinos, SN neutrinos are down-scattered in energy while C$\nu$B neutrinos are up-scattered.  (Hence, the $\nu$SI process does not conserve the number of relativistic, detectable neutrinos, but does conserve the total energy.)  Our constraints on $\nu$SI come from the resulting net dampening of the energies of the SN neutrinos.  Second, upon scattering on the C$\nu$B, SN neutrinos are deflected from their original propagation direction and this delays their arrival at Earth.  Although our constraints on $\nu$SI do not come from detecting this time delay, our treatment of neutrino propagation accounts for the delay implicitly.

We track the propagation of relativistic neutrinos along the radial direction $r$ from the SN to Earth  including the interaction of SN neutrinos on the C$\nu$B and its effect on the neutrino energy and angular distribution.  To obtain our results, we fix the SN distance to $D = 50$~kpc, inspired by the distance to SN 1987A.  In the absence of $\nu$SI, neutrinos from a core-collapse SN at this distance would be detectable.
In the presence of $\nu$SI, we show that they might not.
(To be conservative, we ignore the $\nu$SI process $2 \nu \rightarrow 4 \nu $ during propagation; it is sub-dominant because the center-of-mass energies are significantly lower than in the SN core.  Including it would only intensify the energy dampening of SN neutrinos.)
 
We write the transport equation in terms of the number density of relativistic neutrinos within the energy interval $\Delta E$ and traveling along directions within the interval $\Delta \cos\theta$, $f(E,\cos\theta) \equiv \Delta E ~ \Delta\cos\theta ~ (dn_\nu / dE d\cos\theta)$:
\begin{widetext}
  \begin{eqnarray}
  \frac{d\tf(E,\cos\theta)}{dr} 
  &=&
  \frac{n_{{\rm C}\nu{\rm B}} }{\cos\theta}
  \left\{
  \left. 2 \int_{E}^{\infty} dE_i
  \frac{d\sigma_{2\nu \rightarrow 2\nu}(E_i, \theta-\theta_i)} {dE_f}
  f(E_i,\cos\theta)\right|_{E_{f}=E}
  \right. \nonumber\\
  &-&
  \left. \left .\int_{0}^{E} dE_f
  \frac{d\sigma_{2\nu \rightarrow 2\nu}(E_i, \theta_f-\theta)} {dE_f}
  f(E_{i},\cos\theta)\right|_{E_{i}=E}\right\}\ ,
  \label{equ:evol1}
 \end{eqnarray}
\end{widetext}
where $r$ is the radial distance traveled by neutrinos from the SN and the $\nu$SI cross section $d\sigma_{2\nu \to 2\nu}/dE_f$ is defined in Eq.\ \ref{jac}.
The number density of C$\nu$B neutrinos is $n_{{\rm C}\nu{\rm B}} \approx 330$~cm$^{-3}$. In the case of Dirac neutrinos, half of the neutrinos and antineutrinos will be in sterile states resulting in a reduction by a factor of two of the number of targets. The straight-line direction from the SN to Earth is $\cos \theta = 1$; angular deflections due to scattering are measured relative to that.  The pre-factor $1/\cos\theta$ changes the propagation distance to radial distance; it is approximately equal to unity because the final-state SN neutrino in the interaction is forward-boosted.  
This reflects the fact that, though $\nu$SI may introduce angular deflections in the trajectory of relativistic neutrinos, these are small, so only neutrinos emitted along the direct line-of-sight direction from the SN to Earth, or very close to it, will reach us.


The first term on the right-hand side of Eq.\ \ref{equ:evol1} is a gain term; it accounts for neutrinos of final energy $E$, coming from the scattering of neutrinos that had an initially higher energy. In the gain term, $E_i$ and $\theta_i$ are the initial energy and direction of the neutrinos that, after undergoing $\nu$SI, contribute neutrinos with final energy $E$ and direction $\theta$.  Thus, for given values of $E$ and $\theta$, within the integral, $\theta_i$ is determined by requiring that the final energy is $E_f(E_i, \theta-\theta_i) = E$; see Eq.\ \ref{equ:sigma}.  The pre-factor of 2 in the gain term ensures energy conservation; for each incoming relativistic neutrino, there are two outgoing relativistic neutrinos: the SN neutrino down-scattered in energy and the C$\nu$B neutrino up-scattered in energy.

The second term on the right-hand side of Eq.\ \ref{equ:evol1} is a loss term; it accounts for neutrinos of initial energy $E$ that, after interacting, have final energies $E_f < E$.  In the loss term, $E_i = E$ and $\theta$ are the initial energy and direction of the neutrinos that, after undergoing $\nu$SI, contribute to neutrinos with final energy $E_f < E$ and direction $\theta_f$.  Thus, for given values of $E$ and $\theta$, within the integral, $\theta_f$ is determined by requiring that the final energy is $E_f$.

We evolve $f$ from $r=0$, using the SN neutrino emission as initial condition, to Earth, for a SN at  distance $r = D$.
Because we have assumed that the $\nu$SI coupling is flavor-universal, we neglect flavor conversions in the source and on the way to Earth.

\begin{figure*}[t!]
 \centering
 \includegraphics[width=\columnwidth]{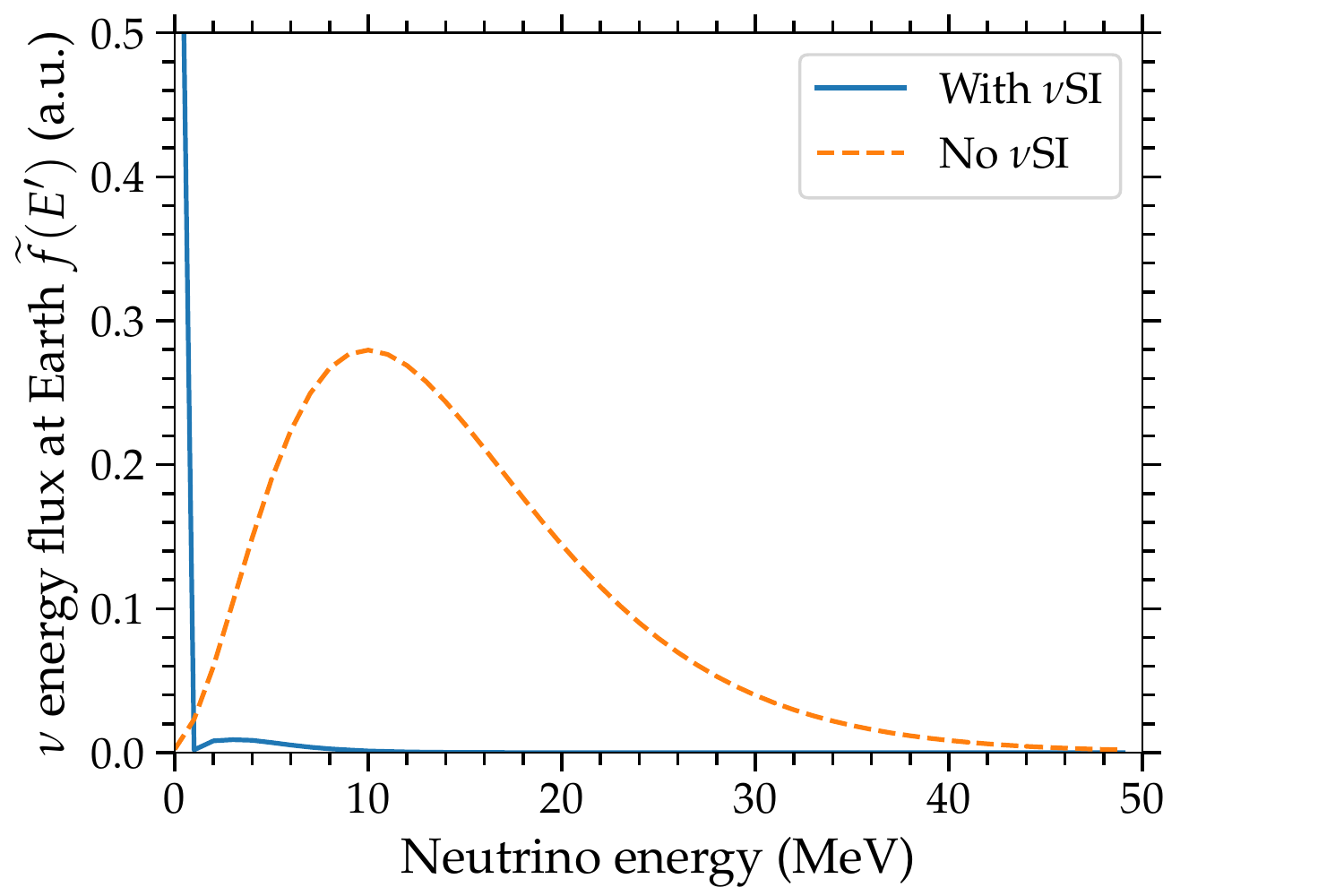}
 \includegraphics[width=\columnwidth]{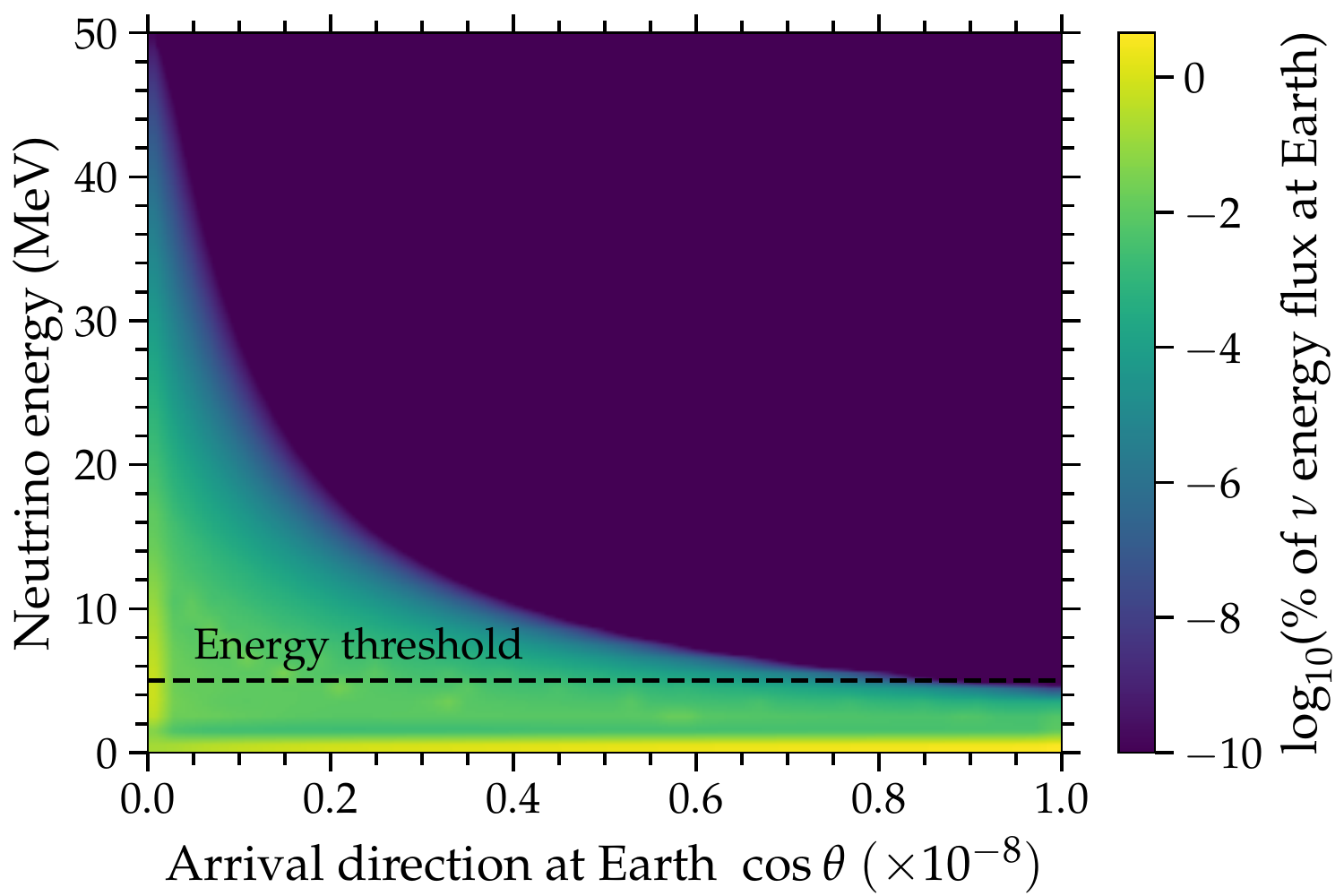}
 \caption{\label{fig:scan_monoenergetic} Illustration of the energy down-scattering of SN neutrinos reaching Earth due to  $\nu$SI  scattering off the C$\nu$B.  {\it Left:} Neutrino energy distribution in arbitrary units before and after $\nu$SI as a function of the neutrino energy for $\log_{10}(g) = -1.75$ and $\log_{10}(M/\textrm{MeV}) = -3.50$. Supernova neutrinos are down-scattered as a result of $\nu$SI. {\it Right:} The color scale represents the  percentage of neutrinos that reach Earth for a SN located 50~kpc away.  The $x$-axis is the arrival direction of the neutrinos at Earth, where $\cos \theta = 1$ means straight-line propagation from the SN to Earth,
 while the $y$-axis is the SN neutrino energy after $\nu$SI on C$\nu$B neutrinos. The initial energy spectrum is same one as  in the left panel. The dashed horizontal line marks the detection energy threshold of $5$~MeV. 
 }
\end{figure*}

To constrain the $\nu$SI parameter space, we solve Eq.~\ref{equ:evol1} for a wide range of values of $M$ and $g$.
For the initial neutrino energy distribution, we use a representative pinched energy distribution~\cite{Keil:2002in, Tamborra:2012ac} with $\langle E \rangle = 15$~MeV and $\langle E^2 \rangle = 290$~MeV$^2$ (see the ``No $\nu$SI'' curve in the left panel of Fig.~\ref{fig:scan_monoenergetic}).  However, the $\nu$SI limits that we obtain are not strongly sensitive to the choice of $\langle E \rangle$ and $\langle E^2 \rangle$. 
We then identify the $(M,g)$ combinations for which the energy down-scattering of SN neutrinos is so intense that the neutrino energy  would fall below a typical energy threshold of detection of 5~MeV~\cite{Scholberg:2017czd}.

Figure~\ref{fig:scan_monoenergetic} illustrates the energy down-scattering of SN neutrinos en route to Earth for the particular choice of $\log_{10}(M/\textrm{MeV}) = -3.50$ and $\log_{10}(g) = -1.75$, and under the assumption that neutrinos are emitted by the SN in a single instantaneous burst.  Supernova neutrinos are scattered below the detection threshold, and the deviation from the radial direction is larger for a larger reduction of neutrino energy.  

To produce our constraints, we repeat this procedure varying the values of $M$ and $g$. Figure~\ref{fig:fig1} shows our constraints on $M$ and $g$ obtained in this way (``SN 1987A propagation'').  Motivated by the detection of neutrinos with 5--40~MeV from SN 1987A, we rule out values of $M$ and $g$ for which at least $99\%$ of all neutrinos are down-scattered below 5~MeV by the time they reach Earth.  Incidentally, our constraints rule out the region of the parameter space recently used to explain the LSND data in Ref.~\cite{Jones:2019tow}.

The cross section for $\nu + \bar{\nu} \rightarrow \nu + \bar{\nu}$ (Eq.~\ref{equ:sigma}) explains the shape of the constrained region in \figu{fig1}.  The resonance energy is $E_{\rm res} \sim 1$~keV.  For masses $M \lesssim 1$~keV, $M$ is negligible compared to the center-of-mass energy $\sqrt{s}$; hence, the constraint in the $\nu$SI parameter space is independent of $M$.  For masses $M \gtrsim 1$~keV, $M$ becomes comparable to or larger than $\sqrt{s}$; hence, the $\nu$SI bounds  weaken with higher masses.  At $M \approx 1$~keV, $M \sim s$ and the cross section is resonant.  Because of multiple scatterings, the resonance is diluted and appears as a shallow dip at $M \approx 1$~keV.

Figure~\ref{fig:fig1} shows that our bounds from propagation cover a smaller region of the parameter space than the bounds from Ref.~\cite{Kolb:1987qy} (``SN 1987A Kolb \& Turner''), which were also derived from propagation.  This is because Ref.~\cite{Kolb:1987qy} assumed a much larger neutrino mass than us, which was allowed at the time.  As a result, the resonance energy $E_{\rm res}$ was much smaller than ours, and bounds for the regime where constraints are independent of $M$ are absent from that work.

Because the typical center-of-mass energy of $\nu$SI inside the SN core is higher than the one  of $\nu$SI during propagation, the $\nu$SI constraints from inside the SN apply to higher values of $M$ and across a wider range.


\section{Conclusions} \label{sec:conclusion} 

Secret neutrino interactions ($\nu$SI) are proposed neutrino-neutrino interactions beyond, and potentially stronger, than the ones foreseen within the Standard Model.    Finding evidence of $\nu$SI or constraining them would provide precious guidance to extend the Standard Model.  We have explored the effect of $\nu$SI on neutrinos from  core-collapse supernovae (SNe).  From the observation of neutrinos from SN 1987A, we have constrained the mass $M$ and coupling strength $g$ of the  new mediator through which $\nu$SI occur.

We have placed two bounds on $\nu$SI: from considering $\nu$SI of SN neutrinos among themselves inside the SN core and from considering $\nu$SI of SN neutrinos that interact with neutrinos from the cosmic neutrino background (C$\nu$B) on their way to Earth.  In deriving these bounds, we have made conservative choices.  Because our methods are largely insensitive to the specific values of the SN parameters adopted as input, our bounds are robust against variations in the neutrino emission across different SNe.

In the SN core, $\nu$SI have the effect of decreasing the overall amount of energy deposited by neutrinos for the shock revival.   As a result, the SN explosion may be halted.  Because the SN 1987A neutrino data support a SN explosion mechanism that is powered by neutrinos, we disfavor the region of the $(M,g)$ parameter space where neutrinos would be unable to achieve this.   

During propagation of SN neutrinos to Earth, $\nu$SI on the C$\nu$B down-scatter the energy of the detectable SN neutrinos.  Because we observed neutrinos with tens of MeV from SN 1987A, we require that $99\%$ of  emitted SN neutrinos reach Earth with energies above a typical detection threshold of 5~MeV, and we disfavor the region of the $(M,g)$ parameter space where this does not occur.  

Our bounds from inside the SN core rely on the next-to-leading-order $\nu$SI process {$2 \nu  \rightarrow 4 \nu $}, whose cross section scales as $g^8$.  The small probability of interaction is outweighed by the extremely large density of neutrinos in the core.  On the other hand, the constraints from the propagation of SN neutrinos to Earth rely on the leading-order $\nu$SI process $\nu + \bar{\nu} \rightarrow \nu + \bar{\nu}$, whose cross section scales as $g^4$.   In this case, the probability of interaction is enhanced by the long distance traveled by neutrinos to Earth.

Figure~\ref{fig:fig1} shows that our bounds from the SN core are more stringent than our bounds from propagation and apply to a wider mediator mass range.  Our bounds are the strongest from $M \approx 10$~MeV to 1~GeV.  For $M \lesssim 100$~MeV, we exclude $g \gtrsim 10^{-1.8}$.  For $M \gtrsim 100$~MeV, our bounds are the first.


\acknowledgments 

We are grateful to George Fuller and Georg Raffelt for insightful discussions. This project was supported by the Villum Foundation (Project No.~13164), the Carlsberg Foundation (CF18-0183), the Knud H\o jgaard Foundation, and the Deutsche Forschungsgemeinschaft through Sonderforschungbereich SFB~1258 ``Neutrinos and Dark Matter in Astro- and Particle Physics'' (NDM).


\appendix*
\medskip
\section{Approximate $2\nu \to 4\nu$ $\nu$SI cross section}
\label{sec:appendix}

\begin{figure*}
\includegraphics[width=0.32\textwidth]{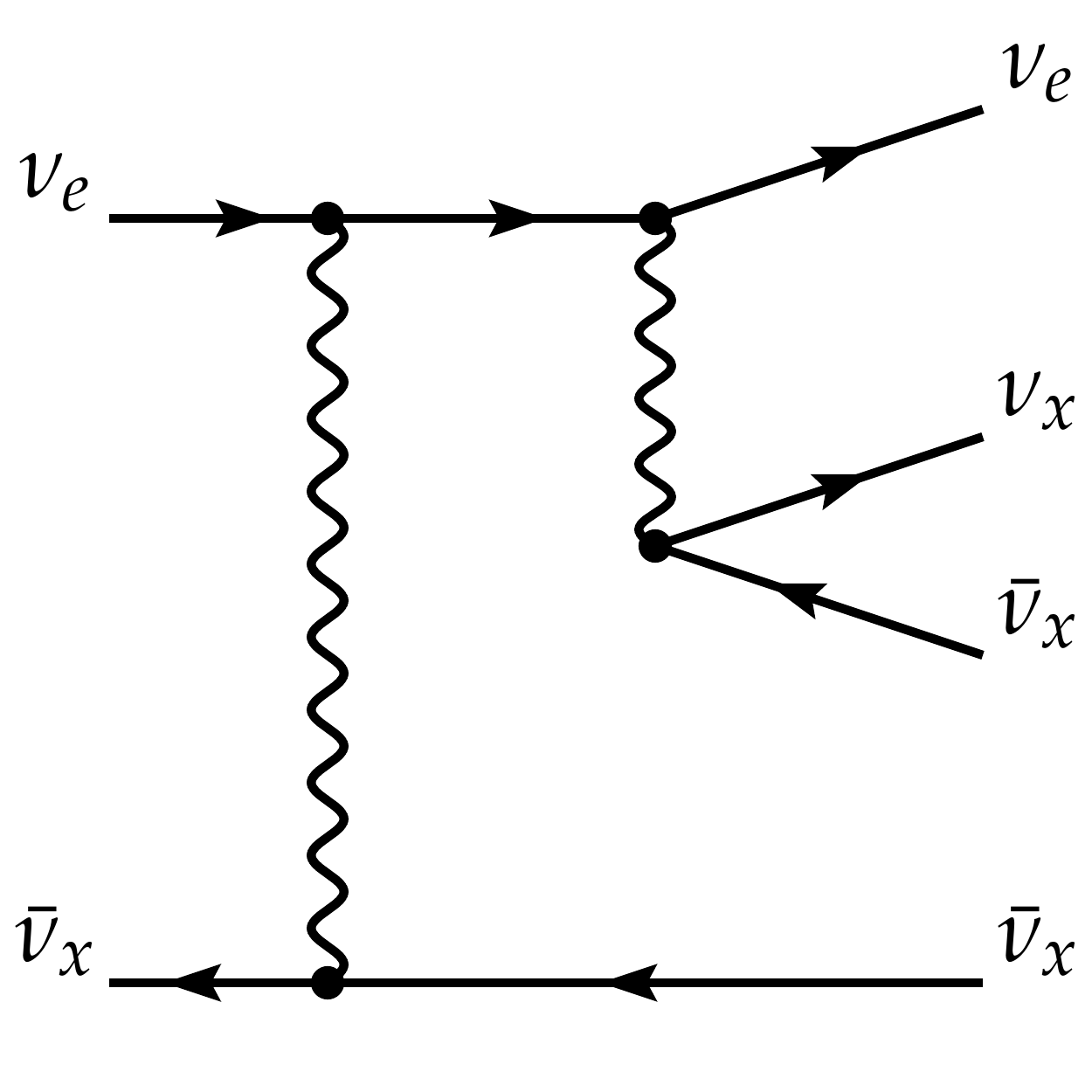}
\includegraphics[width=0.32\textwidth]{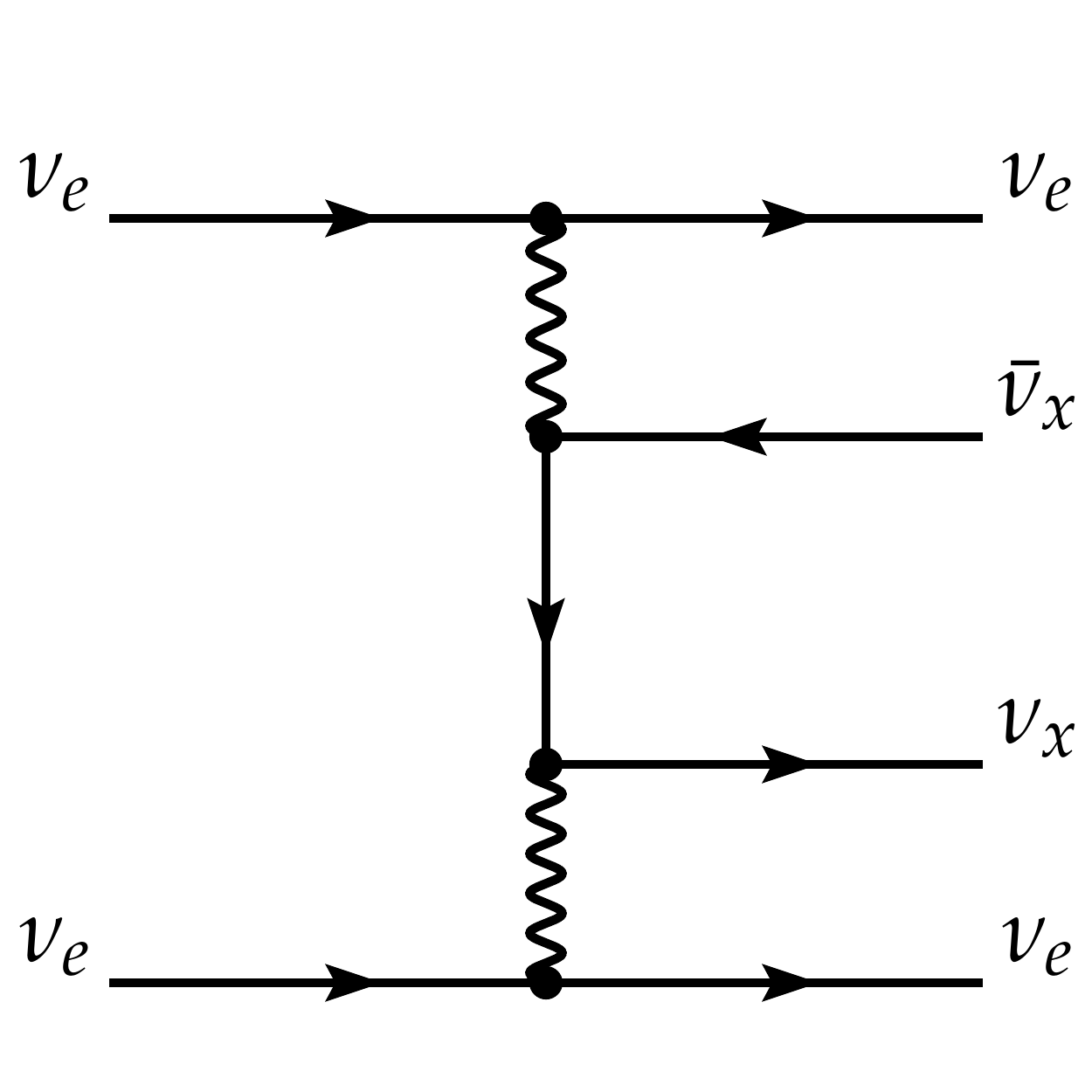}
\includegraphics[width=0.32\textwidth]{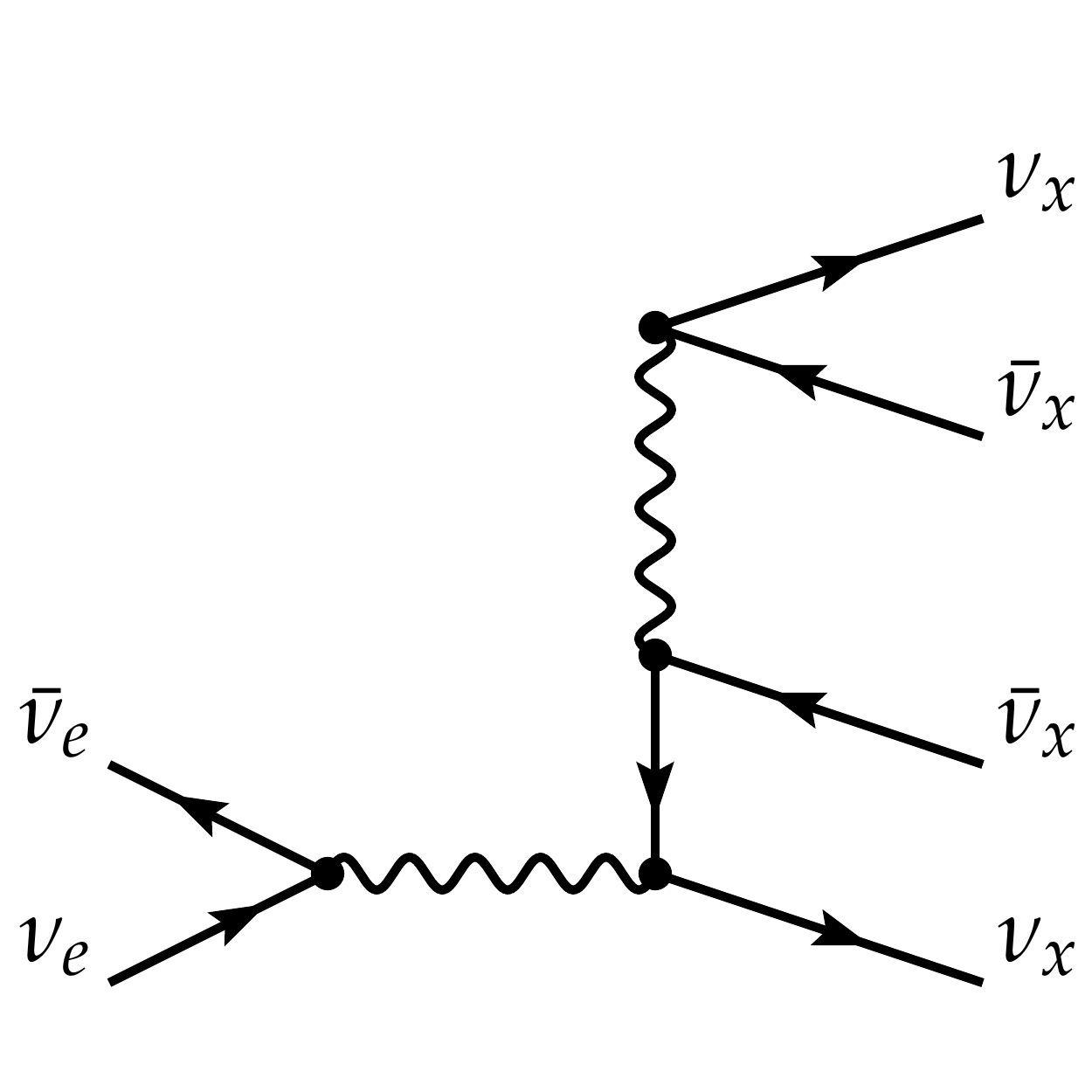}
\caption{Feynman diagrams for some of the illustrative $2\nu \to 4\nu$ $\nu$SI processes.}
\label{feyn}
\end{figure*}

For a $\nu$SI process that converts 2 neutrinos to 4 neutrinos, $2\nu \to 4\nu$, there are more than a hundred Feynman diagrams that contribute to the calculation of the cross section.  Figure~\ref{feyn} shows illustrative topologies of the diagrams; similar diagrams can be obtained by changing the flavors of the neutrinos in the initial and final states.  Because of the large number of contributing diagrams and the interference between them, the precise phase-space integration required to compute the cross section is computationally demanding, and firmly beyond the scope of this work.

\begin{figure}
\includegraphics[width=0.49\textwidth]{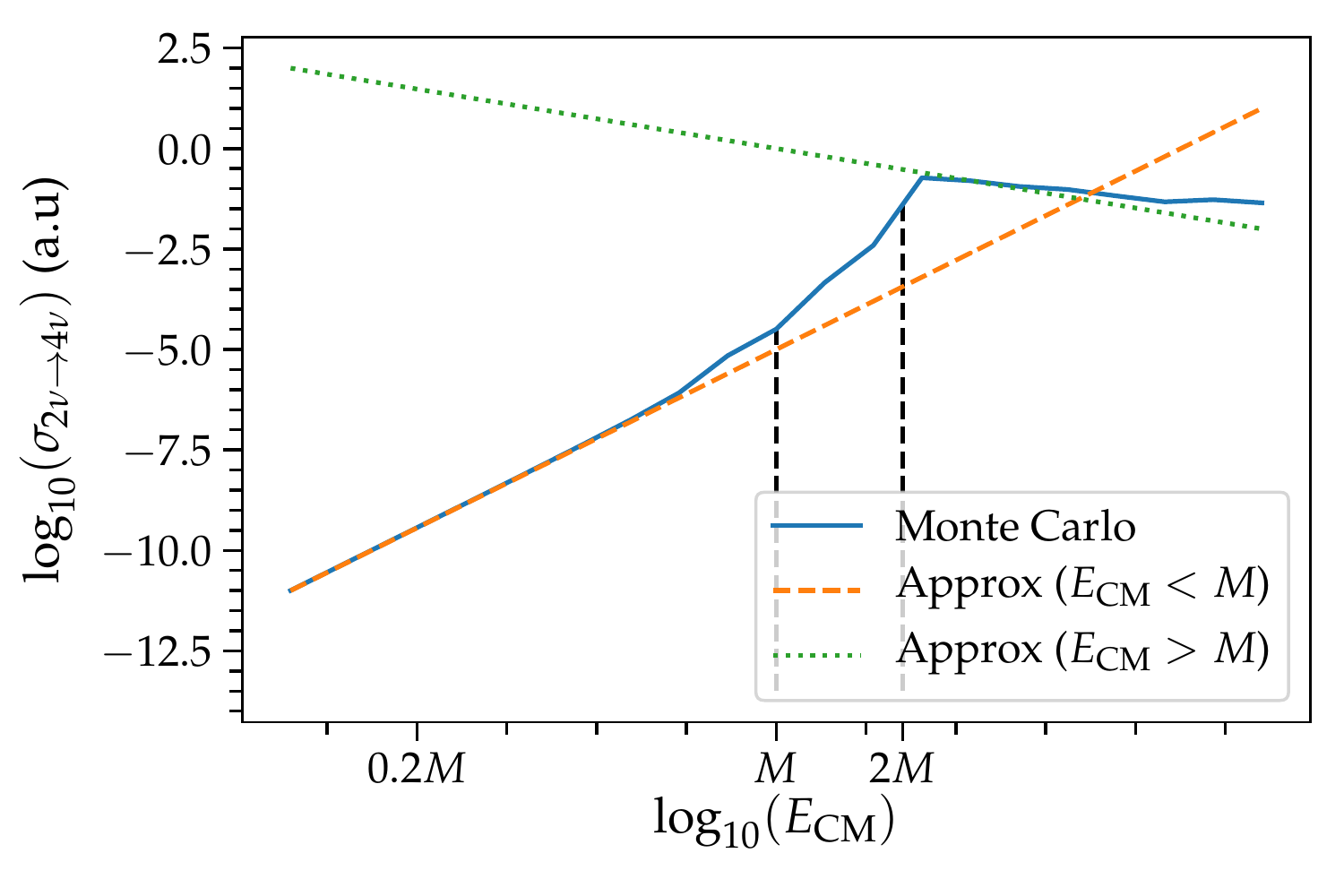}
\caption{Energy dependence of the cross section $\sigma_{2\nu \rightarrow 4\nu}$ $\nu$SI process as a function of the center-of-mass energy, for our approximation, Eqs.~\ref{equ:sigone} and \ref{equ:sigtwo}, and for the precise cross section computed using a single representative contributing diagram.}
\label{xsec}
\end{figure}

We have circumvented such a taxing calculation by using instead the approximate cross section $\sigma_{2\nu\to4\nu}$ in Eqs.~\ref{equ:sigone} and \ref{equ:sigtwo} as an estimate derived solely from dimensional analysis; see the main text for details.  To test the validity of our estimate, we have compared its energy dependence with that of the cross section calculated numerically and precisely, via CalcHEP~\cite{Belyaev:2012qa}, using a single representative contributing Feynman diagram. We have used the first diagram from the left in Fig.~\ref{feyn} for this purpose. Figure~\ref{xsec} shows that their energy dependencies are reasonably similar, confirming the validity of our adopted approximate cross section in our analysis. Because the purpose of this figure is to illustrate the energy dependence of $\sigma_{2\nu\to4\nu}$ regardless of the value of the coupling constant, we use arbitrary units for the cross section.

\bibliography{refs.bib}

\appendix

\end{document}